\documentclass[%
 aip,
 jmp,
 amsmath,amssymb,
 reprint,%
]{revtex4-1}

\usepackage{graphicx}
\usepackage{dcolumn}
\usepackage{bm}

\usepackage[utf8]{inputenc}
\usepackage[T1]{fontenc}
\usepackage{mathptmx}
\usepackage{etoolbox}

\makeatletter
\def\@email#1#2{%
 \endgroup
 \patchcmd{\titleblock@produce}
  {\frontmatter@RRAPformat}
  {\frontmatter@RRAPformat{\produce@RRAP{*#1\href{mailto:#2}{#2}}}\frontmatter@RRAPformat}
  {}{}
}%
\makeatother
\begin{document}

\preprint{AIP/123-QED}

\title{Fully integrated interior solutions of GR for stationary rigidly rotating cylindrical perfect fluids.}

\author{M.-N. C\'el\'erier}
\email{marie-noelle.celerier@obspm.fr}
\affiliation{Laboratoire Univers et Th\'eories, Observatoire de Paris, Universit\'e PSL, Universit\'e Paris Cit\'e, CNRS, F-92190 Meudon, France}

\date{\today}

\begin{abstract}
In an important series of articles published during the 70's, Krasi\'nski displayed a class of interior solutions of the Einstein field equations sourced by a stationary isentropic rotating cylinder of perfect fluid. However, these solutions depend on an unspecified arbitrary function, which lead the author to claim that the equation of state of the fluid could not be obtained directly from the field equations but had to be added by hand. In the present article, we use a double ansatz which we have developed in 2021 and implemented at length into a series of recent papers displaying exact interior solutions for a  stationary rotating cylindrically symmetric fluid with anisotropic pressure. This ansatz allows us to obtain here a fully integrated class of solutions to the Einstein equations, written with the use of very simple analytical functions, and to show that the equation of state of the fluid follows naturally from these field equations.
\end{abstract}


\maketitle 

\section{Introduction} \label{intro}

Finding exact interior solutions for spacetimes sourced by a gravitating fluid is usually considered as a rather involved task. This is the reason why the solutions available in the literature are fewer than those for, e. g.,  vacuum, and why the available ones involve strongly simplifying assumptions such as, e. g., symmetries. In this regards, the case of a stationary cylindrically symmetric perfect fluid rotating around its symmetry axis has been formally solved by Krasi\'nski in the 70's and has been presented then in a series of very interesting papers \cite{K74,K75a,K75b,K78}. However, the most general family of solutions displayed by this author depends on an arbitrary function, denoted there $f$, whose explicit expression has been left unspecified \cite{K78}. It has been claimed, however, that the presence of this arbitrary function $f$ in the solutions hinted at the fact that the equation of state of the fluid could not be obtained directly from the field equations but had to be added by hand.

Now, in a series of recent articles \cite{C21a,C21b,C22a,C22b,C22c}, which will be here referred to as Paper 1-5, respectively, we have developed a powerful method to allow us to integrate the Einstein field equations in the case of a rotating cylinder of nondissipative fluid exhibiting one independent nonvanishing pressure component. This method has been applied successfully to stationary rigidly rotating fluids with different configurations of anisotropic pressure for which a set of classes of exact solutions of General Relativity (GR) has been found. 

In Papers 1 \cite{C21a} and 2 \cite{C22a}, the case where pressure is axially directed has been considered. The special exact solution published in Paper 1 has been developed in Paper 2 where a fully general method allowing the construction of different classes of such solutions has been displayed and exemplified. The investigation of interior spacetimes sourced by such stationary cylindrical anisotropic fluids has been pursued and specialized to an azimuthally directed pressure in Paper 3 \cite{C21b} where another general method for constructing different classes of exact solutions to the field equations adapted to this case has been proposed. Exemplifying such a recipe, a bunch of solutions have been constructed. A number of classes and subclasses have been thus studied and an analysis of their features has lead to the sorting out of two specially interesting classes. In Paper 4 \cite{C22b}, a fluid with radially directed pressure has been considered. Since a generic differential equation, split into three independent parts, has emerged there from the field equations, three corresponding different classes of solutions have been identified. Two of them could only be partially integrated. The other one yielded a set of fully integrated solutions with negative pressure. Physical processes where a negative pressure is encountered have been depicted and gave a rather solid foundation to this class of solutions. For each class described in any of these four papers, axisymmetry and, when appropriate, regularity conditions, matching to an exterior vacuum, proper metric signature, weak and strong energy conditions have been considered while other properties and general rules have been exhibited, some shedding light on rather longstanding issues. A number of astrophysical and physical applications has also been suggested. Finally, in Paper 5 \cite{C22c}, key results and issues raised in the previous works have been synthesized and some fundamental notions of GR, causality, regularity of Lorentzian manifolds, elementary flatness in the vicinity of a symmetry axis, singularities, physics of angular deficits, weak and strong energy conditions have been revisited. Then, a new derivation of the corresponding dust solution has been displayed and has been shown to correspond to the Lanczos-van Stockum solution, shedding new light to this well-known spacetime.

The general methods developed through this series of papers and employed again here are based on the use of two auxiliary functions. The first one, $D$, is expressed as a combination of three among the four metric functions and has been coined, in a simplified design, by van Stockum \cite{vS37} in 1937. The second one, $h$, introduced by the present author in Paper 1 \cite{C21a}, is the ratio of the nonzero pressure component over the energy density of the fluid. Using this double ansatz, we construct here a class of solutions for the interior spacetimes of a stationary rigidly rotating cylindrically symmetric perfect fluid, directly issued from the field equations, and show that the equation of state follows naturally from these field equations. Moreover, the expressions for the metric functions and the physical properties of the fluid imply simple analytical functions very easy to deal with for any possible future application or theoretical purpose. Finally, the couple of constant parameters determining each solution in this class, is given a straightforward physical interpretation. This improves the results obtained by Krasi\'nski and shows that his statement about the independence of the equation of state is not fully general. We must stress however that our class of solutions, although being a genuine analytic solution to the field equations of GR without any extra assumption added, represents only part of the solutions to the problem since it is issued from merely one among the two independent differential equations yielded by these field equations. Since the second one is not solvable without addition of arbitrary assumption(s), we suggest that it might be, as displayed in our formalism, the generator of the remaining solutions in the Krasi\'nski class.

Our paper is organised as follows. In Sec. \ref{sp}, the problem is posed and the field and resulting basic equations to be solved are provided. In Sec. \ref{solver}, two different classes of solutions are identified through two independent differential equations issued from the field equations. In Sec. \ref{sol} the metrics belonging to the first class are solved. In Sec \ref{ic}, a number of constraints on the parameters of the solutions are discussed and implemented when needed, while the main physical and mathematical features of these solutions are calculated and analyzed. In Sec. \ref{final}, the final two-parameter form of this class of solutions and the corresponding equation of state of the fluid are displayed. Our conclusions are provided in Sec. \ref{concl}.

\section{Description of the problem} \label{sp}

The gravitational source is here a stationary cylindrically symmetric perfect fluid rigidly rotating around its axis. It is bounded by a cylindrical hypersurface $\Sigma$. Its isotropic pressure $P$ is left unspecified, i. e., no equation of state is a priori imposed. Its stress-energy tensor can therefore be written as
\begin{equation}
T_{\alpha \beta} = (\rho + P) V_{\alpha}V_{\beta} + P g_{\alpha \beta}, \label{r1}
\end{equation}
where $\rho$ denotes the energy density of the fluid and $V_\alpha$, its timelike 4-velocity, that satisfies
\begin{equation}
V^{\alpha}V_{\alpha}=-1. \label{r2}
\end{equation}

As for the previous spacetimes of Papers 1-5, a spacelike hypersurface orthogonal Killing vector $\partial_z$ is assumed, such as to ease a subsequent proper junction to a vacuum exterior Lewis metric. Therefore, in geometric units $c=G=1$, the line element reads
\begin{equation}
\textrm{d}s^2=-f \textrm{d}t^2 + 2 k \textrm{d}t \textrm{d}\phi +\textrm{e}^\mu (\textrm{d}r^2 +\textrm{d}z^2) + l \textrm{d}\phi^2. \label{metric}
\end{equation}
The metric coefficients $f$, $k$, $\mu$, and $l$ are assumed to be real functions of the radial coordinate $r$ only, such as to account for stationarity. The cylindrical symmetry imposes that the coordinates conform to the following ranges:
\begin{equation}
- \infty \leq t \leq +\infty, \quad 0 \leq r \leq +\infty, \quad -\infty \leq z \leq +\infty \quad 0 \leq \phi \leq 2 \pi, \label{ranges}
\end{equation}
the two limits of $\phi$, $0$ and $2\pi$, being topologically identified. The coordinates are denoted $x^0=t$, $x^1=r$, $x^2=z$, and $x^3=\phi$.

Rigid rotation allows the choice of a frame corotating with the fluid \cite{CS20,C21a,D06}. Thus, its 4-velocity can be written as
\begin{equation}
V^\alpha = v \delta^\alpha_0, \label{r3}
\end{equation}
with $v$ a function of $r$ only. Therefore, the timelike condition for $V^\alpha$ displayed in (\ref{r2}) becomes
\begin{equation}
fv^2 = 1. \label{timeliker}
\end{equation}
To allow the integration of the field equations, we introduce the ansatz used in Papers 1-5, i. e., two key auxiliary functions of the radial coordinate $r$, $D(r)$ written as \cite{D06}
\begin{equation}
D^2 = fl + k^2, \label{D2}
\end{equation}
and the $h(r)$ function \cite{C21a} defined as
\begin{equation}
h=\frac{P}{\rho}. \label{hdef}    
\end{equation}

\subsection{Field equations} \label{fe}

Inserting (\ref{r3})--(\ref{timeliker}) into (\ref{r1}), we obtain the components of the stress-energy tensor matching the nonvanishing components of the Einstein tensor. Then, using (\ref{D2}), the five corresponding field equations can be written as

\begin{eqnarray}
G_{00} &=& \frac{\textrm{e}^{-\mu}}{2} \left[-f\mu'' - 2f\frac{D''}{D} + f'' - f'\frac{D'}{D} + \frac{3f(f' l' + k'^2)}{2D^2}\right] \nonumber \\
&=& \kappa\rho f, \label{G00r}
\end{eqnarray}
\begin{eqnarray}
G_{03} &=&  \frac{\textrm{e}^{-\mu}}{2} \left[k\mu'' + 2 k \frac{D''}{D} -k'' + k'\frac{D'}{D} - \frac{3k(f' l' + k'^2)}{2D^2}\right] \nonumber \\
&=& - \kappa\rho k, \label{G03r}
\end{eqnarray}
\begin{equation} 
G_{11} = \frac{\mu' D'}{2D} + \frac{f' l' + k'^2}{4D^2} = \kappa P \textrm{e}^{\mu}, \label{G11r}
\end{equation}
\begin{equation}
G_{22} = \frac{D''}{D} -\frac{\mu' D'}{2D} - \frac{f'l' + k'^2}{4D^2} =\kappa P \textrm{e}^{\mu} , \label{G22r}
\end{equation}
\begin{eqnarray}
G_{33} &=&  \frac{\textrm{e}^{-\mu}}{2} \left[l\mu'' + 2l\frac{D''}{D} - l'' + l'\frac{D'}{D} - \frac{3l(f' l' + k'^2)}{2D^2}\right] \nonumber \\
&=&  \frac{\kappa}{f}\left(\rho k^2 + P D^2 \right), \label{G33r}
\end{eqnarray}
where the primes stand for differentiation with respect to $r$.

\subsection{Conservation of the stress-energy tensor} \label{bi}

The conservation of the stress-energy tensor is implemented by the Bianchi identity, whose general form is available as (57) in C\'el\'erier and Santos \cite{CS20}. Specialized to the present case, it becomes
\begin{equation}
T^\beta_{1;\beta} = P' + (\rho + P) \frac{f'}{2f} = 0. \label{Bianchi1r}
\end{equation}
With $h(r)$ as defined in (\ref{hdef}), it reads
\begin{equation}
\frac{P'}{P} + \frac{(1+h)}{2h} \frac{f'}{f} = 0. \label{Bianchi2r}
\end{equation}

\subsection{Some reminders} \label{rem}

The main key equations established in previous works and still applying here are recalled below. That initially displayed as (14) in Debbasch et al. \cite{D06} is written here as
\begin{equation}
kf' - fk' = 2c D, \label{r6}
\end{equation}
where $c$ is an integration constant and the factor 2 is added for further convenience. This first-order ordinary differential equation in $k$ possesses, as a solution, see (24) of Paper 1,
\begin{equation}
k = f \left(c_k - 2c \int^r_0 \frac{D(v)}{f^2(v) } \textrm{d} v \right). \label{r7}
\end{equation}

Another useful relation, established as (19) in Paper 2, is
\begin{equation}
\frac{f'l' +k'^2}{2D^2} = \frac{f'D'}{fD} - \frac{f'^2}{2f^2} + \frac{2c^2}{f^2}. \label{r8}
\end{equation}

\section{Identifying two classes of solutions} \label{solver}

Using another scheme analogous to that displayed in Paper 4 to allow the integration of the Bianchi identity, another auxiliary function $F(r)$ is defined by
\begin{equation}
\frac{F'}{F} = (1+h) \frac{f'}{f}. \label{r9}
\end{equation}
Inserted into Bianchi's identity (\ref{Bianchi2r}), it gives
\begin{equation}
\frac{P'}{P} + \frac{F'}{2F} = 0, \label{r10}
\end{equation}
which can be integrated as
\begin{equation}
P F^{\frac{1}{2}} = c_B, \label{r11}
\end{equation}
where $c_B$ is an integration constant.

Now, adding (\ref{G11r}) and (\ref{G22r}), we obtain
\begin{equation}
\frac{D''}{D} = 2 \kappa P \textrm{e}^{\mu}. \label{r12}
\end{equation}
Then, subtracting (\ref{G11r}) from (\ref{G22r}) gives
\begin{equation}
\frac{f'l' +k'^2}{2D^2} = \frac{D''}{D} - \frac{\mu'D'}{D}, \label{r12a}
\end{equation}
which is inserted into (\ref{G00r}) together with (\ref{r12}) to give
\begin{equation}
\left(3\mu' + \frac{f'}{f} \right) \frac{D'}{D} = -\mu'' -\frac{(1-h)}{h}\frac{D''}{D}  + \frac{f''}{f}. \label{r12b}
\end{equation}
 
Now, $P$ given by (\ref{r11}) is inserted into (\ref{r12}) to yield
\begin{equation}
\frac{D''}{D} = 2\kappa c_B\frac{\textrm{e}^{\mu}}{F^{\frac{1}{2}}}. \label{r13}
\end{equation}
Then, (\ref{r13}) is inserted into (\ref{G22r}), together with (\ref{r8}) and (\ref{r11}), to give
\begin{equation}
\left(\mu' + \frac{f'}{f} \right) \frac{D}{D}' = \frac{f'^2}{2f^2} - \frac{2c^2}{f^2}  + \frac{2 \kappa c_B \textrm{e}^{\mu}}{F^{\frac{1}{2}}}. \label{r14}
\end{equation}
Then, (\ref{r13}) is inserted into (\ref{r12b}), which yields
\begin{equation}
\left(3\mu' + \frac{f'}{f} \right) \frac{D'}{D} = -\frac{2(1-h)}{h}\kappa c_B\frac{\textrm{e}^{\mu}}{F^{\frac{1}{2}}}  -\mu''  + \frac{f''}{f}. \label{r12c}
\end{equation}
Equalizing both expressions for $D'/D$ as given by (\ref{r14}) and (\ref{r12c}), we obtain
\begin{equation}
\frac{\kappa c_B \textrm{e}^{\mu}}{F^{\frac{1}{2}}} = - \frac{\left[\left(\mu''-\frac{f''}{f}\right)\left(\mu'+\frac{f'}{f}\right) + \left(3\mu'+\frac{f'}{f}\right)\left(\frac{f'^2}{2f^2} -\frac{2c^2}{f^2}\right)
\right]}{4\mu'+\frac{2}{h}\left(\mu'+\frac{f'}{f}\right)}.  \label{r14a}
\end{equation}

Now, inserting (\ref{r11}) into (\ref{r14}), then differentiating the result with respect to $r$ and using the expression for $P'$ as given by (\ref{Bianchi2r}), we have
\begin{eqnarray}
&& \left(\mu'' + \frac{f''}{f} - \frac{f'^2}{f^2}\right)\frac{D'}{D} + \left(\mu'+\frac{f'}{f}\right)\left(\frac{D''}{D}-\frac{D'^2}{D^2}\right)-\frac{f'f''}{f^2} \nonumber \\
&+& \frac{f'^3}{f^3} - \frac{4c^2f'}{f^3}
= \kappa P \textrm{e}^{\mu}\left[2\mu' - \frac{(1+h)}{h}\frac{f'}{f}\right],  \label{r15}
\end{eqnarray}
where $D''/D$ is replaced by its expression (\ref{r12}), $P$ by its expression issued from (\ref{r11}), then $D'/D$ by that extracted from (\ref{r14}), and $\kappa c_B \textrm{e}^{\mu}/F^{1/2}$ by that given by (\ref{r14a}) which yields
\begin{eqnarray}
&& \left(\mu' + \frac{f'}{f}\right)\left\{\left(\mu' + \frac{f'}{f}\right)\left[\left(\frac{1-h}{h}\right)\left(\frac{f'^2}{2f^2} - \frac{2c^2}{f^2}\right) \right. \right. \nonumber \\
&-& \left. \left.\left(\mu''-\frac{f''}{f}\right)\right]^2   
- \left[\mu''+\frac{f''}{f} + \left(\frac{1+h}{2h}\right) \frac{f'^2}{f^2} + \left(\frac{1+3h}{2h}\right)\frac{\mu'f'}{f} \right] \right. \nonumber \\
&\times& \left. \left[\left(\frac{1+2h}{h}\right)\mu' + \frac{f'}{hf}\right] \left[\left(\frac{1-h}{h}\right) \left(\frac{f'^2}{2f^2} - \frac{2c^2}{f^2}\right) -\left(\mu''-\frac{f''}{f}\right)\right] \right. \nonumber \\
&+& \left. \frac{f'}{f}\left[\left(\frac{1-h}{2h}\right)\left(\frac{f'^2}{2f^2} - \frac{2c^2}{f^2}\right) + \frac{f''}{f}\right] \right. \nonumber \\
&\times& \left. \left[\left(\frac{1+2h}{h}\right)\mu' + \frac{f'}{hf}\right]^2 \right\} = 0. \label{r16}
\end{eqnarray}

\section{Solving the field equations} \label{sol}

The left hand side of (\ref{r16}) is the product of two factors. As in Paper 4, the extra degree of freedom, issuing from the fact that the five field equations do not match the number of the six unknowns ($f$, $k$, $\textrm{e}^{\mu}$, $l$, $\rho$ and $P$), is used in choosing to cancel out one or the other factor. The second one yields a very complicated second order in both $\mu$ and $f$ differential equation and it has not been possible to simplify it nor to use it as a seed for an exact analytical solution. The first one instead can generate a fully integrated class of solutions as shown below. The corresponding differential equation is
\begin{equation}
\mu' + \frac{f'}{f} = 0, \label{r17}
\end{equation}
which can be integrated as
\begin{equation}
\textrm{e}^{\mu}=\frac{c_{\mu}}{f}, \label{r18}
\end{equation}
where $c_{\mu}$ is an integration constant.

Inserting (\ref{r17}) into (\ref{r12c}) gives
\begin{equation}
-\mu'' + \frac{f''}{f} -\frac{2 \kappa c_B(1-h)}{h} \frac{\textrm{e}^{\mu}}{F^{\frac{1}{2}}} + \frac{2f'D'}{fD} = 0. \label{r19}
\end{equation}
Then, inserting (\ref{r17}) into (\ref{r14}) we obtain
\begin{equation}
\frac{f'^2}{2f^2} - \frac{2c^2}{f^2} = - \frac{2 \kappa c_B \textrm{e}^{\mu}}{F^{\frac{1}{2}}}. \label{r20}
\end{equation}
Now, (\ref{r17}) is differentiated with respect to $r$ such as to give
\begin{equation}
\mu'' = - \frac{f''}{f} + \frac{f'^2}{f^2}, \label{r21}
\end{equation}
which is itself inserted into (\ref{r19}), together with (\ref{r20}), such as to yield
\begin{equation}
\frac{2f''}{f} + \left(\frac{1-3h}{2h}\right)\frac{f'^2}{f^2} - \left(\frac{1-h}{h}\right) \frac{2c^2}{f^2}+ \frac{2f'D'}{fD} = 0. \label{r22}
\end{equation}

Now, inserting (\ref{r11}), (\ref{r17}) and (\ref{r18}) into (\ref{r14}) gives
\begin{equation}
2\kappa c_{\mu} \frac{P}{f} = -\frac{f'^2}{2f^2}+ \frac{2c^2}{f^2}. \label{r23}
\end{equation}
Then, (\ref{Bianchi2r}) and (\ref{r11}) are inserted into (\ref{r23}) after it has been differentiated with respect to $r$, which yields
\begin{equation}
\left(\frac{1+3h}{h}\right)\frac{\kappa c_B c_{\mu} f}{F^{\frac{1}{2}}} - ff''+f'^2-4c^2 = 0. \label{r24}
\end{equation}
Now, (\ref{r18}) inserted into (\ref{r20}) gives
\begin{equation}
\frac{\kappa c_B c_{\mu}}{F^{\frac{1}{2}}} = -\frac{f'^2}{4f}+ \frac{c^2}{f}, \label{r25}
\end{equation}
that is inserted into (\ref{r24}) to yield
\begin{equation}
\frac{f''}{f} = \left(\frac{1-h}{4h}\right)\left(\frac{4c^2}{f^2} - \frac{f'^2}{f^2}\right), \label{r26}
\end{equation}
which is inserted into (\ref{r22}) such as to obtain
\begin{equation}
\frac{D'}{D} = \frac{f'}{2f}, \label{r27}
\end{equation} 
which can be integrated as
\begin{equation}
D = c_D \sqrt{f}, \label{r28}
\end{equation}  
where $c_D$ is an integration constant.

Now, (\ref{r20}) is inserted into (\ref{r13}) to give
\begin{equation}
\frac{D''}{D} = \frac{2c^2}{f^2} - \frac{f'^2}{2f^2}. \label{r29}
\end{equation} 
Then, the well-known identity
\begin{equation}
\frac{D''}{D} = \left(\frac{D'}{D}\right)' + \frac{D'^2}{D^2}, \label{r30}
\end{equation}
with (\ref{r27}) inserted, becomes
\begin{equation}
\frac{D''}{D} = \frac{f''}{2f} - \frac{f'^2}{4f^2}, \label{r31}
\end{equation}
which is equalized to (\ref{r30}) and gives
\begin{equation}
\frac{f''}{f} = \frac{4c^2}{f^2} - \frac{f'^2}{2f^2}, \label{r32}
\end{equation}
which is equalized to (\ref{r26}) to yield
\begin{equation}
f'^2 = 4c^2 \left(\frac{1-5h}{1-3h}\right), \label{r32a}
\end{equation}
that gives
\begin{equation}
f' = 2 \epsilon c \sqrt{\frac{1-5h}{1-3h}}, \label{r33}
\end{equation}
with $\epsilon = \pm 1$ determining two subclasses of solutions, and which is differentiated with respect to $r$ such as to yield
\begin{equation}
f'' = - 2 \epsilon c \frac{h'}{(1-3h)^{\frac{3}{2}}(1-5h)^{\frac{1}{2}}}. \label{r34}
\end{equation}
Then, (\ref{r33}) and (\ref{r34}) are inserted into (\ref{r32}), which gives
\begin{equation}
f = - \epsilon c \frac{(1-h)(1-3h)^{\frac{1}{2}}(1-5h)^{\frac{1}{2}}}{h'}. \label{r35}
\end{equation}
Dividing (\ref{r33}) by (\ref{r35}), we obtain
\begin{equation}
\frac{f'}{f} = - \frac{2h'}{(1-h)(1-3h)}, \label{r36}
\end{equation}
which can be integrated as
\begin{equation}
f = c_f \frac{(1-3h)}{(1-h)}, \label{r37}
\end{equation}
where $c_f$ is an integration constant.

Inserted into (\ref{r18}), (\ref{r37}) yields
\begin{equation}
\textrm{e}^{\mu} = \frac{c_{\mu}}{c_f} \frac{(1-h)}{(1-3h)}. \label{r38}
\end{equation}
Then, (\ref{r37}) is inserted into (\ref{r28}) such as to give
\begin{equation}
D = c_D \sqrt{\frac{c_f(1-3h)}{(1-h)}}. \label{r39}
\end{equation}

Now, (\ref{r37}) is inserted into (\ref{r35}) such as to yield
\begin{equation}
h' = - \frac{\epsilon c}{c_f} \frac{(1-h)^2 (1-5h)^{\frac{1}{2}}}{(1-3h)^{\frac{1}{2}}}, \label{r40}
\end{equation}
which can be written as an equation for the $r$ coordinate that reads
\begin{equation}
r = - \frac{\epsilon c_f}{c} \int^h_{h_0} \frac{(1-3v)^{\frac{1}{2}}}{(1-v)^2 (1-5v)^{\frac{1}{2}}}  \textrm{d} v , \label{r41}
\end{equation}
where $h_0$ is the value of the ratio $h$ at the axis of symmetry and which becomes, after integration,
\begin{eqnarray}
r &=& \frac{\epsilon c_f}{4 \sqrt{2} c} \left[\tanh^{-1}\left(\sqrt{\frac{1-5h}{2(1-3h)}}\right) + \frac{\sqrt{2(1-3h)(1-5h)}}{(1-h)} \right.\nonumber \\
&-& \left. \tanh^{-1}\left(\sqrt{\frac{1-5h_0}{2(1-3h_0)}}\right) - \frac{\sqrt{2(1-3h_0)(1-5h_0)}}{(1-h_0)} \right]. \label{r42}
\end{eqnarray}

Then, we insert (\ref{r37}) and (\ref{r39}) into (\ref{r7}) and obtain
\begin{equation}
k = c_f \frac{(1-3h)}{(1-h)} \left[c_k - \frac{2c c_D}{c_f^{\frac{3}{2}}}\int^r_0 \frac{(1-v)^{\frac{3}{2}}}{(1-3v)^{\frac{3}{2}}} \textrm{d} v \right], \label{r43}
\end{equation}
where we make a change of variable, while using (\ref{r40}), that gives
\begin{equation}
k = c_f \frac{(1-3h)}{(1-h)} \left[c_k + \frac{2 \epsilon c_D}{\sqrt{c_f}}\int^h_{h_0} \frac{\textrm{d} v} {(1-v)^{\frac{1}{2}} (1-3v)  (1-5v)^{\frac{1}{2}}}  \right], \label{r44}
\end{equation}
which can be integrated as
\begin{eqnarray}
k &=& c_f \frac{(1-3h)}{(1-h)} \left\{c_k - \frac{2 \epsilon c_D}{\sqrt{c_f}} \left[\tan^{-1}\left(\sqrt{\frac{1-5h}{1-h}}\right) \right. \right. \nonumber \\ 
&-& \left. \left. \tan^{-1}\left(\sqrt{\frac{1-5h_0}{1-h_0}}\right) \right] \right\}. \label{r45}
\end{eqnarray}

Now, $l$ is obtained as usual by implementing the definition of the auxiliary function $D$ given by (\ref{D2}) that yields
\begin{eqnarray}
l &=& c_D^2 - c_f \frac{(1-3h)}{(1-h)} \left\{c_k - \frac{2 \epsilon c_D}{\sqrt{c_f}} \left[\tan^{-1}\left(\sqrt{\frac{1-5h}{1-h}}\right) \right. \right. \nonumber \\ 
&-& \left. \left. \tan^{-1}\left(\sqrt{\frac{1-5h_0}{1-h_0}}\right) \right] \right\}^2. \label{r46}
\end{eqnarray}

Finally, substituting (\ref{r36}) into (\ref{Bianchi2r}), we obtain
\begin{equation}
\frac{P'}{P} = \frac{(1+h)h'}{h(1-h)(1-3h)}, \label{r47}
\end{equation}
which can be integrated as
\begin{equation}
P = c_P \frac{h(1-h)}{(1-3h)^2}, \label{r48}
\end{equation}
where $c_P$ is another integration constant and that gives, using the definition of the $h$ ratio,
\begin{equation}
\rho = c_P \frac{(1-h)}{(1-3h)^2}. \label{r49}
\end{equation}

Now, (\ref{r29}) substituted into (\ref{r12}) yields
\begin{equation}
\frac{2c^2}{f^2} - \frac{f'^2}{ 2 f^2} = 2 \kappa P \textrm{e}^{\mu}, \label{r50}
\end{equation}
where (\ref{r36})-(\ref{r38}), (\ref{r40}) and (\ref{r48}) are inserted to obtain
\begin{equation}
c_P = \frac{2c^2}{\kappa c_f c_{\mu}}, \label{r51}
\end{equation}
which can be substituted into (\ref{r48}) and (\ref{r49}) to obtain
\begin{equation}
P =  \frac{2c^2}{\kappa c_f c_{\mu}} \frac{h(1-h)}{(1-3h)^2} \label{r52}
\end{equation}
and
\begin{equation}
\rho = \frac{2c^2}{\kappa c_f c_{\mu}} \frac{(1-h)}{(1-3h)^2}. \label{r53}
\end{equation}

\section{Constraints on the integration constants} \label{ic}

At this stage, our class of solutions depends on six integration constants, $c$, $c_f$, $c_{\mu}$, $c_D$, $c_k$ and $h_0$ and on a sign indeterminacy $\epsilon$. However, mathematical and physical conditions have to be fulfilled by this class and this imposes constraints on these parameters such as to reduce their number to two. Such conditions have already been analyzed and discussed in the previous companion papers, in particular in Paper 5. They are thus implemented below.

\subsection{Rescaling of the coordinates}

Owing to the expression for $\textrm{e}^{\mu}$ given by (\ref{r38}), it is always possible to rescale the coordinates $r$ and $z$ from a factor $c_{\mu}/c_f$, which amounts to setting everywhere
\begin{equation}
c_{\mu} = c_f. \label{res1}
\end{equation}

\subsection{Axisymmetry condition} \label{axi}

The metric function $l$ of an axisymmetric spacetime must vanish in the limit at the axis \cite{M93,P96,S09}, which implies a constraint on the integration constants written as
\begin{equation}
l \stackrel{0}{=} 0, \label{rax1}
\end{equation} 
where $\stackrel{0}{=}$ denotes that the involved quantities are evaluated at the axis where $r=0$. With (\ref{r46}) inserted, this becomes
\begin{equation}
\frac{c_D^2}{c_f c_k} = \frac{1-3h_0}{1-h_0}. \label{rax2}
\end{equation}

From the definition of $D^2$ given by (\ref{D2}), (\ref{rax1}) yields another constraint that reads
\begin{equation}
D^2 \stackrel{0}{=} k^2, \label{rax3}
\end{equation}
where (\ref{r39}) and (\ref{r45}) are substituted which gives
\begin{equation}
\frac{c_f c_k^2}{c_D^2} = \frac{1-h_0}{1-3h_0}, \label{rax4}
\end{equation}
which is inserted into (\ref{rax2}) such as to yield
\begin{equation}
c_k= 1, \label{rax5}
\end{equation}
with which, (\ref{rax2}) becomes
\begin{equation}
\frac{c_D^2}{c_f} = \frac{1-3h_0}{1-h_0}. \label{rax6}
\end{equation}

\subsection{Hydrodynamical tensors, vectors and scalars}

These quantities have been calculated for every class of solutions displayed in previous Papers 1-5. Besides providing interesting properties of the gravitating fluid, an additional constraint on the integration constants can be derived from the theorem using the rotation scalar and provided in Appendix A of Paper 2.

The non-vanishing component of the acceleration vector reads, from (23) of C\'el\'erier and Santos \cite{CS20},
\begin{equation}
\dot{V}_1 = \frac{1}{2} \frac{f'}{f}, \label{rdotV1a}
\end{equation}
which becomes, with (\ref{r33}) and (\ref{r37}) inserted,
\begin{equation}
\dot{V}_1 = \frac{\epsilon c}{c_f} \frac{(1-h)(1-5h)^{\frac{1}{2}}}  {(1-3h)^{\frac{3}{2}}}. \label{rdotV1b}
\end{equation}
Its modulus follows therefore as
\begin{equation}
\dot{V}^\alpha \dot{V}_\alpha = \frac{c^2}{c_f^2} \frac{(1-h)(1-5h)}{(1-3h)^2}. \label{rdotV1c}
\end{equation}

Now, the nonvanishing component of the rotation tensor reads, from  (98) of Paper 4,
\begin{equation}
2 \omega_{13} = \frac{2 c D}{f^{\frac{3}{2}}}, \label{romega13b}
\end{equation}
where we insert (\ref{r37}) and (\ref{r39}) to obtain
\begin{equation}
\omega_{13} = \frac{c c_D}{c_f} \left(\frac{1-h}{1-3h} \right). \label{romega13c}
\end{equation}
The rotation scalar follows from (86) of Paper 1 as
\begin{equation}
\omega^2 = \frac{c^2}{f^2\textrm{e}^{\mu}}, \label{romega2a}
\end{equation}
which becomes, after inserting (\ref{r37}) and (\ref{r38}) and implementing (\ref{res1})
\begin{equation}
\omega^2 = \frac{c^2}{c_f^2} \left(\frac{1-h}{1-3h} \right). \label{romega2b}
\end{equation}

Here, the theorem demonstrated in Appendix A of Paper 2 can be used. It states that any class of solutions such as those studied here, verifying (\ref{r6}) together with an equation of the form of Paper 2's (A1), which corresponds here to (\ref{r17}), satisfies
\begin{equation}
\omega^2 \stackrel{0}{=} c^2, \label{romega2c}
\end{equation}
which, inserted into (\ref{romega2b}) written at the axis, yields
\begin{equation}
c_f^2 = \frac{1-h_0}{1-3h_0}. \label{romega2d}
\end{equation}

As it is well-known, the shear tensor vanishes for any rigidly rotating fluid \cite{D06,C21a}.

\subsection{Real nature of the solutions}

For an easy matching to the Weyl class of the Lewis vacuum exterior in view of future applications to standard physical systems, we limit our study to real number solutions.

First, notice that $\sqrt{c_f}$ appears in the expressions for $k$, $l$ and $D$, implying therefore $c_f>0$. The same reasoning applies to the expression $\sqrt{(1-5h)/(1-h)}$ appearing in $k$ and $l$. Hence, $1-5h$ and $1-h$ must exhibit the same sign, whatever the value of $h$. This implies
\begin{equation}
\textrm{either} \quad h < \frac{1}{5} \quad \quad \textrm{or} \quad 1<h \qquad \forall h. \label{sign1}
\end{equation}

\subsection{Metric signature}

As stressed in Papers 1-5, an actual mathematical solution of the field equations is not obligatorily a genuine spacetime of GR. A key feature which has to be realized is the Lorentzian character of the metric signature. Owing to the form (\ref{metric}) retained here for the line element, the four metric functions need therefore to be all positive or all negative definite.

Now, considering $f$ and $\textrm{e}^{\mu}$ this implies $c_f$ and $c_{\mu}$ both positive since $c_f$ has already been required to be positive. This is consistent with the constraint (\ref{res1}).

Then, the same sign must be displayed by $f$ and $k$ whatever the value of $h$ and, in particular, for $h=h_0$. This implies $c_k>0$.

Moreover, $c_f>0$ and $c_k>0$ inserted into (\ref{rax2}) give
\begin{equation}
\frac{1-3h_0}{1-h_0}>0, \label{sign2}
\end{equation}
which is perfectly consistent with (\ref{sign1}) and implies $f$ and $\textrm{e}^{\mu}$ positive definite at the axis, since $c_f$ and $c_{\mu}$ are equal and positive. Thus, $f$ and $\textrm{e}^{\mu}$ must be everywhere positive, the same applying to $k$ that gives
\begin{equation}
\frac{1-3h}{1-h}>0 \quad \forall h. \label{sign3}
\end{equation}

Now, from (\ref{r45}), we can write
\begin{equation}
k \stackrel{0}{=} c_f c_k \left(\frac{1-3h_0}{1-h_0} \right), \label{sign4}
\end{equation}
which, substituted into (\ref{rax2}), yields
\begin{equation}
k \stackrel{0}{=} c_D^2, \label{sign5}
\end{equation}
that implies $k$ positive at the axis. Hence, $k$ being a smooth function of $r$, we can always choose the value of the radial coordinate on the boundary, $r_{\Sigma}$, such that $k(r_{\Sigma})>0$ and $k$ positive definite everywhere inside the cylinder.

Now, the sign of the metric function $l$ is analyzed as follows. Since, owing to the axisymmetry condition, $l$ vanishes on the axis, and since an examination of (\ref{r46}) shows that it does not vanish for any other value of $h$, it suffices that it be increasing from the axis to be positive everywhere. Therefore, its first derivative with respect to $r$ must be itself positive, if only at the axis. Differentiating (\ref{r46}) with respect to $r$ and inserting the expression for $h'$ given by (\ref{r40}), we obtain
\begin{equation}
l' = -2 \epsilon c \sqrt{\frac{1-5h}{1-3h}} \{  \quad \}^2 + \frac{4 c c_D}{\sqrt{c_f}}\sqrt{\frac{1-h}{1-3h}} \{ \quad  \}, \label{sign6}
\end{equation}
where $\{ \quad \}$ is short for the expression inside the brackets in (\ref{r46}). At the axis, this expression reads $\{ \quad \}\stackrel{0}{=} c_k=1$. Therefore, with this value inserted together with
\begin{equation}
\frac{c_D}{\sqrt{c_f}} = \eta \sqrt{\frac{1-3h_0}{1-h_0}}, \label{sign7}
\end{equation}
issued from (\ref{rax2}) and where $\eta = \pm1$, (\ref{sign6}) yields
\begin{equation}
l' \stackrel{0}{=} 2 c \left(2\eta - \epsilon \sqrt{\frac{1-5h_0}{1-3h_0}}\right). \label{sign8}
\end{equation}
A straightforward analysis of the different possible combinations of the signs of $c$, $\epsilon$ and $\eta$, not reproduced here owing to its length, shows that the four combinations for which $l'$ is positive at the axis are:

. $c>0$, $\epsilon>0$ and $\eta>0$,

. $c<0$, $\epsilon<0$ and $\eta<0$,

both for either $h_0<1/5$ or $h_0>1$, and

. $c>0$, $\epsilon<0$ and $\eta>0$,

. $c<0$, $\epsilon>0$ and $\eta<0$,

both whatever the value of $h_0$.

\subsection{Junction condition} \label{junc}

For astrophysical purpose, the interior solutions displayed here are matched to an exterior vacuum. This vacuum is chosen, as it has been justified in Papers 1-4, to be the Weyl class of the Lewis solutions for a stationary rotating cylindrically symmetric vacuum \cite{L32}.

It has been shown by Israel \cite{I66} that a proper matching to such an exterior vacuum spacetime is obtained provided the radial pressure vanishes on the boundary $\Sigma$. In the present case of an isotropic pressure, this yields $P_{\Sigma} = 0$, $P_{\Sigma}$ being the value of the isotropic pressure $P$ on the boundary.

Since $h$ cannot reach the unity, for which the pressure would also vanish, the value of $h$ on the boundary is $h_{\Sigma}=0$ which verifies the constraint $h_{\Sigma}<1/5$.

\subsection{Energy condition} \label{ec}

Different energy conditions can be found in the literature. For an application to standard astrophysical objects, the most generic is the weak energy condition implying $\rho>0$. With $\rho$ given by (\ref{r53}), this constraint implies $h<1$, which, together with (\ref{sign1}), gives
\begin{equation}
h< \frac{1}{5}, \label{sign9}
\end{equation}
that is indeed consistent with the junction condition.

Another interesting constraint proceeds from the strong energy condition which, for a perfect fluid, reads
\begin{equation}
\rho + P > 0, \qquad \rho + 3 P > 0. \label{strong1}  
\end{equation}
With $P$ given by (\ref{r52}) and $\rho$ by (\ref{r53}), the first inequality implies
\begin{equation}
(1-h)(1+h) > 0, \label{strong2}  
\end{equation}
and the second inequality yields
\begin{equation}
(1-h)(1+3h) > 0. \label{strong3}   
\end{equation}
Satisfying both inequalities implies therefore
\begin{equation}
- \frac{1}{3} < h < 1. \label{strong4} 
\end{equation}
Coupled with (\ref{sign1}), this constraint finally reads
\begin{equation}
- \frac{1}{3} < h < \frac{1}{5}, \label{strong5} 
\end{equation}
which is also consistent with the junction condition.

\subsection{Behaviour of the $h(r)$ function} \label{hr}

The derivative of $h(r)$ given by (\ref{r40}) vanishes obviously for $h=1/5$ or $h=1$.

Now, $h=1$ is never reached owing to the weak energy condition (\ref{sign9}). Therefore, $h'$ can possibly vanish at the limiting value of $h$, $h=1/5$.

At the axis, the derivative of $h$ reads
\begin{equation}
h' \stackrel{0}{=} - \frac{\epsilon c}{c_f} \frac{(1-h_0)^2 (1-5h_0)^{\frac{1}{2}}}{(1-3h_0)^{\frac{1}{2}}}. \label{sign10}
\end{equation}
Hence, $h'$ exhibits everywhere the same sign as $- \epsilon c$.

i) case where $- \epsilon c>0$

In this case, $h'>0$ and $h$ should be a function of $r$ increasing from $h_0$: $0<h_0<1/5$ to $h_{\Sigma}=0$ which is of course impossible. Case i) is therefore ruled out.

ii) case where $- \epsilon c<0$

In this case, $h'<0$ and $h$ is a function of $r$ decreasing from $h_0$: $0<h_0<1/5$ to $h_{\Sigma}=0$ that summarizes as
\begin{equation}
\frac{1}{5}>h_0>h>h_{\Sigma}=0. \label{sign12}
\end{equation}
Hence, the pressure is positive definite, which is generic for standard applications in astrophysics.

\subsection{Regularity (elementary flatness) condition} \label{regcond}

The constraints described above are more or less mandatory.The metric signature, the axisymmetry condition and other constraints issuing from the particular features of the solutions, such as those imposed by the rotation scalar theorem and the positiveness of the square root radicals must obligatorily be fulfilled by the solutions so that hey can be considered as actual real GR solutions of the problem.

On the other hand, the junction and weak energy conditions, even if necessary in a standard astrophysical context, are not fundamental to validate the solutions as genuine GR ones.

The status of the so-called ''regularity conditions'' is even more complicated. They deal with the issue of avoiding singularities on the axis. Indeed, a method for identifying the existence of hidden singularities on the axis has been proposed by Mars and Senovilla \cite{M93} and developed further on \cite{C00,S09}. It is linked to the notion of angular completeness and angular deficit and has been discussed at length in Paper 5. Based on the requirement of elementary flatness in the vicinity of the rotation axis, it consists in checking whether the ratio of the circumference over the radius of an infinitesimally small circle around the axis departs or not from $2 \pi$. The circle describes the orbit of the spacelike Killing vector $\vec{\xi}$ generating the azimuthal isometry. This Killing vector should therefore satisfy the so-called ''regularity'' condition \cite{S09}, which, for cylindrical symmetry and the coordinate frame retained here reads
\begin{equation}
\frac{\textrm{e}^{-\mu}l'^2}{4l} \stackrel{0}{=} 1, \label{sign13}
\end{equation}
where (\ref{r37}), (\ref{r38}) and (\ref{r40}) are inserted to give
\begin{equation}
\frac{c^2}{c_f}\left(2 c_D \sqrt{1-h_0} -\epsilon\sqrt{c_f(1-5h_0)}\right)^2 = c_D^2(1-h_0)-c_f(1-3h_0), \label{sign14}
\end{equation}
which implies
\begin{equation}
c_D^2 > c_f \frac{(1-3h_0)}{1-h_0}. \label{sign15}
\end{equation}

Now, as it has been stated by Wilson and Clarke \cite{W96}, this so-called ''regularity'' or ''elementary flatness'' condition does not ensure the smoothness of the manifold on the axis. This statement has been confirmed by the analysis, in Paper 5, of this condition applied to the different solutions displayed in Papers 1-5.

It is therefore given here merely for completeness in order to be applied wisely to possible relevant configurations. However, any solution which would not satisfy it might anyhow be considered as a proper GR solution, its relevance depending, of course, on the intended application. 

\subsection{Summary of the constraints on the integration constants} \label{sic}

The constraints on the constant parameters are summarized here then combined, the result being that the solutions are finally depending on only two independent parameters.

The rescaling of the coordinates $r$ and $z$ imposes $c_{\mu} = c_f$ as given by (\ref{res1})

The axisymmetry condition is displayed in (\ref{rax2}), implying a relation between $c_D$, $c_f$ , $c_k$ and $h_0$.

The definition of $D^2$ yields $c_k=1$, provided in (\ref{rax5}).

The rotation scalar theorem gives $c_f$ as an expression implying only $h_0$ and given by (\ref{romega2d}).

Finally, the ''regularity condition'' is implemented by (\ref{sign14}) that links $c$, $c_D$ $c_f$ and $h_0$.

Now, combining (\ref{rax2}), (\ref{rax5}) and (\ref{romega2d}), we obtain
\begin{equation}
c_D^2 = \sqrt{\frac{1-3h_0}{1-h_0}}. \label{sign16}
\end{equation}
Then, (\ref{romega2d}) and (\ref{sign16}) inserted into the ''regularity condition'' (\ref{sign14}) yield
\begin{equation}
h_0 = \frac{3}{7}, \label{sign17}
\end{equation}
that is incompatible with the weak energy condition, $h<1/5$.

Therefore, if we choose to stick to the weak energy condition and get rid of the so-called ''regularity condition'', we are left with the remaining constraints on the integration constants 
\begin{equation}
c_k = 1, \label{sign18}
\end{equation}
\begin{equation}
c_{\mu} = c_f = \sqrt{\frac{1-h_0}{1-3h_0}}, \label{sign19}
\end{equation}
\begin{equation}
c_D^2 = \sqrt{\frac{1-3h_0}{1-h_0}}. \label{sign20}
\end{equation}

\section{Final form of the solutions and equation of state} \label{final}

\subsection{Final solutions}

The above relations (\ref{sign18})--(\ref{sign20}) between the parameters are now substituted into the expressions describing the solutions that become
\begin{equation}
f = \sqrt{\frac{1-h_0}{1-3h_0}} \frac{(1-3h)}{(1-h)}, \label{f1}
\end{equation}
\begin{equation}
\textrm{e}^{\mu} = \frac{1-h}{1-3h}, \label{f2}
\end{equation}
\begin{eqnarray}
k &=& \sqrt{\frac{1-h_0}{1-3h_0}} \frac{(1-3h)}{(1-h)} \left\{1 - 2 \epsilon \sqrt{\frac{1-3h_0}{1-h_0}} \left[\tan^{-1}\left(\sqrt{\frac{1-5h}{1-h}}\right) \right. \right. \nonumber \\ 
&-& \left. \left. \tan^{-1}\left(\sqrt{\frac{1-5h_0}{1-h_0}}\right) \right] \right\}, \label{f3}
\end{eqnarray}
\begin{eqnarray}
l &=& \sqrt{\frac{1-3h_0}{1-h_0}} - \sqrt{\frac{1-h_0}{1-3h_0}} \frac{(1-3h)}{(1-h)} \left\{1 - 2 \epsilon \sqrt{\frac{1-3h_0}{1-h_0}} \right. \nonumber \\  
&\times& \left.\left[\tan^{-1}\left(\sqrt{\frac{1-5h}{1-h}}\right) - \tan^{-1}\left(\sqrt{\frac{1-5h_0}{1-h_0}}\right) \right] \right\}^2, \label{f4}
\end{eqnarray}
\begin{equation}
D = \sqrt{\frac{1-3h}{1-h}}, \label{f5}
\end{equation}
\begin{equation}
h' = - c \sqrt{\frac{1-3h_0}{1-h_0}}\frac{(1-h)^2 \sqrt{1-5h}}{\sqrt{1-3h}}, \label{f6}
\end{equation}
\begin{eqnarray}
r &=& \frac{1}{4 \sqrt{2} c}\sqrt{\frac{1-h_0}{1-3h_0}} \left[\tanh^{-1}\left(\sqrt{\frac{1-5h}{2(1-3h)}}\right)  \right.\nonumber \\
&+& \left. \frac{\sqrt{2(1-3h)(1-5h)}}{1-h}
-  \tanh^{-1}\left(\sqrt{\frac{1-5h_0}{2(1-3h_0)}}\right) \right. \nonumber \\
&-& \left.\frac{\sqrt{2(1-3h_0)(1-5h_0)}}{1-h_0} \right], \label{f7}
\end{eqnarray}
\begin{equation}
P = \frac{2 c^2}{\kappa} \frac{(1-3h_0)}{(1-h_0)}\frac{h(1-h)}{(1-3h)^2}, \label{f8}
\end{equation}
\begin{equation}
\rho = \frac{2 c^2}{\kappa} \frac{(1-3h_0)}{(1-h_0)} \frac{(1-h)}{(1-3h)^2}. \label{f9}
\end{equation}
These solutions depend therefore on two independent parameters, here chosen to be two quantities with a robust physical interpretation: the amplitude of the rotation on the axis, $c$, and the ratio of the pressure over the energy density on the axis, $h_0$. Since $\epsilon c$ must be positive, as shown in Sec. \ref{hr}, we have set $\epsilon = 1$ in (\ref{f6}) and (\ref{f7}), adopting therefore the convention $c>0$.

\subsection{Singularities} \label{rsing}

Owing to their above expressions, the four metric functions, the energy density and the radial pressure are diverging or vanishing only for $h$ or $h_0$ equal to $1/3$ or to $1$, both values situated outside the interval $[0,1/5]$ allowed to this ratio. Therefore, the corresponding spacetimes are singularity-free.

\subsection{Equation of state} \label{eos}

The results obtained here imply that the equations of state possibly verified by the fluid are imposed by the field equations. Indeed, five independent field equations have been used to determine six unknowns, the four metric functions, the energy density and the pressure. However, the remaining degree of freedom have been used to choose to set to vanish one among the two factors in (\ref{r16}). Thus, a class of solutions involving exact expressions for the pressure and the density has been obtained, which means that the equation of state of the corresponding fluid is determined as shown below.

Inserting (\ref{hdef}) into (\ref{f9}), one obtains a second degree equation in $P$ that reads
\begin{equation}
9P^2 + 2 \left[\frac{c^2}{\kappa} \frac{(1-3h_0)}{(1-h_0)} -3\rho \right] P + \rho\left[\rho -\frac{2 c^2}{\kappa} \frac{(1-3h_0)}{(1-h_0)} \right] =0, \label{eos1}
\end{equation}
whose solutions are
\begin{equation}
P = \frac{\rho}{3} - \frac{c^2}{9\kappa} \frac{(1-3h_0)}{(1-h_0)} \left[1 - \eta \sqrt{1 + \frac{12 \kappa (1-h_0) \rho}{c^2 (1-3h_0)}} \right], \label{eos2}
\end{equation}
with $\eta = \pm 1$. Since the pressure $P$ given by (\ref{eos2}) is indeed positive, as imposed by the constraints discussed in Sec. \ref{hr}, whatever the sign of $\eta$, we are left with two possible equations of state.

These equations of state depart from that of an ultra-relativistic gas with isotropic pressure, i. e., $P= \rho /3$, by a quantity
\begin{equation}
\Delta = - \frac{c^2}{9\kappa} \frac{(1-3h_0)}{(1-h_0)} \left[1 - \eta \sqrt{1 + \frac{12 \kappa (1-h_0) \rho}{c^2 (1-3h_0)}} \right]. \label{eos3}
\end{equation}
This quantity can vanish only in the case when $\eta=+1$ and provided $\rho$ should also vanish that would imply an equation of state for dust, which is not the case considered here. Thus the equations of state (\ref{eos2}), directly derived from the field equations, apply as such and the amplitude of their departure from the polytropic case depends essentially on the respective values of $c$ and $h_0$ that determine each particular solution in the class. It is, however, interesting to note that the smaller $c$, i. e., the amplitude of the rotation at the axis, the closer the equation of state to the polytropic one for an ultra-relativistic gas. This is a new counter-intuitive result provided by the GR framework.

\subsection{Agreement with the Krasi\'nski solutions}

Two particular features pertaining both to the system considered by Krasi\'nski and to the present configuration, but introduced differently here and there, are shown now to be equivalent so that both problems end up being indeed strictly identical.

One of the key assumptions made by Krasi\'nski \cite{K74} is that the fluid is subjected to isentropic motion, which is characterized by the condition which, in our notations, reads
\begin{equation}
\rho = \rho(P). \label{ag1}
\end{equation}
This condition is actually realized by the equation of state (\ref{eos1}) which is indeed analogous to implying an isentropic motion of the fluid studied in the present article.

A second statement made by Krasi\'nski \cite{K74} is that the shear of the velocity field of the fluid studied there vanishes. This property implies that the fluid is rigidly rotating which is also a feature of the system assumed here.

All the other explicit assumptions being the same in both studies, the corresponding problems are rigorously analogous and the class of solutions displayed here can be considered as a subclass of Krasi\'nski's, whose fully analytic expressions (\ref{f1})-(\ref{f9}) might allow an easier use for possible future applications and for a better understanding of GR.

\section{Conclusions} \label{concl}

We have considered the interior spacetimes sourced by a stationary rigidly rotating cylinder of perfect fluid with nonvanishing isotropic pressure. The solutions to the Einstein field equations exhibited here are exact, fully integrated, two parameter dependent and displayed in a form allowing a very straightforward physical interpretation. They are therefore ready for applications in a number of cases, elongated astrophysical objects rotating around their axis of symmetry, topological defects, (super)-strings, etc. 

Elongated astrophysical objects are numerous in the Universe. Such are interstellar columns of gas and astrophysical jets, exhibiting much more extended lengths, that are driven from diverse objects on very different size and mass scales. These can emerge from the vicinity of supermassive black holes in the case of active galactic nuclei (AGN), of star-sized black holes in the case of microquasars, of neutron stars making up X-ray binaries, from protostellar cores in young stellar objects, and from white dwarfs in symbiotic binaries and supersoft X-ray sources. At cosmological scales, the filaments making up the cosmic web might also be modeled by some infinite cylinders of matter whose first approximation could involve a perfect fluid.

At the other end of the length scale, topological defects with the shape of cosmic string or super-strings are objects whose width is so small that they are usually studied in the zero-width approximation. Therefore, an infinite cylinder might represent them accurately, and the present class of solutions might be of use provided a perfect fluid approximation should apply.

As usual in GR, and, more generally, in physics, the use of exact solutions to represent actual phenomenons implies some approximations. However, new exact solutions of the field equations allow one to make progress in the understanding of gravitation from both a mathematical and physical point of view.

The solutions exhibited here have been obtained thanks to the essential use of two auxiliary functions and various calculation methods developed in a series of five preceding papers named Papers 1-5 \cite{C21a,C21b,C22a,C22b,C22c}. However, they do not cover the whole set of solutions to the problem, since they have been obtained by using only one of the two differential equations issued from the field equations. The class of spacetimes corresponding to the second equation remains therefore to be found, presumably with the use of a numerical method. 

We have shown that the problem considered here is exactly the same as that studied by Krasi\'nski in the 70's \cite{K74,K75a,K75b,K78} and partially solved by this author under the name ''first family''. Indeed the class of solutions to the field equations displayed there depends on an unspecified function $f$. Since the solutions found  by Krasi\'nski for this issue seem to be exhaustive, while not fully determined, our second differential equations might be considered as being the generator of the remaining, not fully integrated, solutions of the Krasi\'nski first family.

The results can be therefore stated as: Krasi\'nski family I solutions being exhaustive but not fully integrated, those displayed here represent the fully integrable part of this family.

Another outstanding result is that, to each solution pertaining to the here displayed class, corresponds a couple of equations of state for the fluid, issued directly from the field equations. The only degree of freedom left here consists of the choice between both expressions. Hence, the equation of state cannot be imposed by hand at will, contrary to what is usually done when dealing with the stress-energy tensor of a fluid in GR. Its departure from that of an ultra-relativistic gas depends on the values of the two parameters defining each solution in the class and decreases with the rotation velocity.

As it has been done for the exact solutions displayed in the preceding series of Papers 1-5, the main mathematical and physical properties of the present perfect fluid exact solutions have been analyzed and discussed at length. Imposing the weak energy condition implies that the energy density and the pressure of the fluid are both positive definite and that no singularity spoils the described spacetimes. 

We have shown that the strong energy condition can also be satisfied by these solutions for an additional constraint on the allowed values of the pressure over energy density ratio.

Moreover, these solutions have been properly matched to an exterior Lewis vacuum. Therefore, these exact perfect fluid solutions can be of use for a number of standard or less standard applications to sufficiently elongated rotating systems where shear can be neglected.

\appendix
\section{Short reminder of the steps to obtain the main previously derived equations boldly used here} \label{omega}

In Papers 1-5, a number of key equations have been derived and some have been used as such in the present main text. For completeness purpose, we give here a short reminder of the main steps to obtain these equations.

As recalled in Sec. \ref{rem}, (\ref{r7}) is the general solution of the first order ordinary differential equation in $k$ (\ref{r6}), as one can easily verify by a mere substitution of this solution into (\ref{r6}).

The relation (\ref{r8}) is issued from inserting and arranging (\ref{D2}) into (\ref{r6}).

Equation (\ref{romega13b}) proceeds from the use of (\ref{r6}) into (115) of C\'el\'erier and Santos \cite{CS20}.

Finally, using (\ref{romega13b}), the rotation scalar $\omega$, defined by
\begin{equation}
\omega^2 = \frac{1}{2} \omega^{\alpha \beta}\omega_{\alpha \beta}, \label{omega2def}
\end{equation}
follows as
\begin{equation}
\omega^2 = \frac{1}{4  \textrm{e}^{\mu}D^2}\left(k\frac{f'}{f}-k'\right)^2. \label{omega2a}
\end{equation}
Inserting (\ref{r6}) into (\ref{omega2a}), one obtains
(\ref{romega2a}).

\end{document}